\documentclass[12pt]{article}

\usepackage{amsmath,amsfonts,amssymb,latexsym}

\newtheorem{theorem}{Theorem}[section]
\newenvironment{proof}[1][Proof]{\textsc{#1.} }{\ \rule{0.5em}{0.5em}}
\numberwithin{equation}{section}

\begin{document}
\title{{\Huge Global Hyperbolicity of Sliced Spaces}}
\author{{\Large Spiros Cotsakis}\\D.I.C.S.E. \\ University of the Aegean\\
Karlovassi 83 200, Samos, Greece\\ \texttt{email: skot@aegean.gr}}
\maketitle
\begin{abstract}
\noindent We show that for generic sliced spacetimes global
hyperbolicity is equivalent to space completeness under the
assumption that the lapse, shift and spatial metric are uniformly
bounded. This leads us to the conclusion that simple sliced spaces
are timelike and null geodesically complete if and only if space
is a complete Riemannian manifold.
\end{abstract}

\section{Introduction}
\noindent The singularity theorems of general relativity inform us
that the singular behaviour met in the simplest isotropic and
homogeneous but anisotropic spacetimes is a more general
phenomenon: It arises in all those circumstances which, under
canonical causality assumptions such as global hyperbolicity,
globally share the same geodesic properties (geodesic focusing) as
that found in the simplest models.

However, this by no means exhausts the available possibilities
even in the simplest examples. For instance, the geometric
properties of the universal covering space of AdS space are
refreshingly unlike those of the usual FRW spaces. AdS space is
not globally hyperbolic but it is (null and spacelike)
geodesically complete.

The purpose of this paper is to discuss a connection between
global hyperbolicity and space completeness in spacetimes which
are very general, but are closer in a sense to AdS behaviour
rather than that of the usual FRW universes and their
generalizations. In a simple sub-class of such spacetimes, all
three, geodesic completeness, slice completeness and global
hyperbolicity, are equivalent.

\section{Slice completeness and global hyperbolicity}
Consider a spacetime of the form $(\mathcal{V},g)$ with
$\mathcal{V}=\mathcal{M} \times \mathcal{I},\;$ $\mathcal{I}$
being an interval in $\mathbb{R}$ and $\mathcal{M}$  a smooth
manifold of dimension $n$, in which the smooth,
$(n+1)$--dimensional, Lorentzian metric $g$ splits as follows:
\begin{equation}
g\equiv -N^{2}(\theta ^{0})^{2}+g_{ij}\;\theta ^{i}\theta
^{j},\quad \theta ^{0}=dt,\quad \theta ^{i}\equiv dx^{i}+\beta
^{i}dt.  \label{2.1}
\end{equation}
Here $N=N(t,x^{i})$ is called the \emph{lapse function}, $\beta
^{i}(t,x^{j})$ is called the \emph{shift function} and the spatial
slices $\mathcal{M}_{t}\,(=\mathcal{M}\times \{t\})$ are spacelike
submanifolds endowed with the time-dependent spatial metric
$g_{t}\equiv g_{ij}dx^{i}dx^{j}$. We call such a spacetime a
\emph{sliced space} \cite{cb49}.

Sliced spaces are time-oriented by increasing $t$ and may be
thought of as warped products $\mathcal{M}\times_{N}\mathbb{R},\,
N:\mathcal{M}\rightarrow\mathbb{R}$ with the lapse as their
warping function (in general the lapse is defined in the extended
space $\mathcal{M}\times\mathbb{R}$). Notice, however, that such a
warped product is different than the usual warped product form
$\mathbb{R}\times_{f}\mathcal{M},\,f:\mathbb{R}\rightarrow\mathbb{R}$,
which includes, for instance, the \textsc{FRW} spaces. The
simplest example of a sliced space which cannot be written in the
usual warped product form is the (universal covering space of) AdS
spacetime.

For simplicity we choose $\mathcal{I}=\mathbb{R}$. The following
hypothesis insures that the parameter $t$ measures, up to a
positive factor bounded above and below, the proper time along the
normals to the slices $\mathcal{M}_{t}.$ We say that a sliced
space has \emph{uniformly bounded lapse} if the lapse function $N$
is bounded below and above by positive numbers $N_{m}$ and
$N_{M}$,
\begin{equation}
0<N_{m}\leq N\leq N_{M}. \label{h1}
\end{equation}

A  sliced space has \emph{uniformly bounded shift} if the $g_{t}$
norm of the shift vector $\beta$, projection on the tangent space
to $\mathcal{M}_{t}$ of the tangent to the lines $\{x\}\times
\mathcal{I}$, is uniformly bounded by a number $B.$

A  sliced space has \emph{uniformly bounded spatial metric} if the
time-dependent metric $g_{t}\equiv g_{ij}dx^{i}dx^{j}$ is
uniformly bounded below and above for all $t\in \mathcal{I}$ by a
metric $\gamma =g_{0}$, that is there exist  numbers $A, D>0$ such
that for all tangent vectors $v$ to $\mathcal{M}$ it holds that
\begin{equation}
A\gamma _{ij}v^{i}v^{j}\leq g_{ij}v^{i}v^{j}\leq D\gamma
_{ij}v^{i}v^{j}.  \label{h2}
\end{equation}

 We prove the following result.
\begin{theorem}
Let  $(\mathcal{V},g)$ be a sliced space with uniformly bounded
lapse $N$, shift $\beta$ and spatial metric $g_{t}$. Then the
following are equivalent:
\begin{enumerate}
\item $(\mathcal{M}_{0},\gamma )$ is a complete Riemannian manifold
\item The spacetime $(\mathcal{V},g)$ is globally hyperbolic
\end{enumerate}
\end{theorem}
\begin{proof}
$(1)\Rightarrow (2)$. This was proved in \cite{ch-co02}, but we
present here a somewhat different, but completely equivalent,
proof which is based on Penrose's definition of global
hyperbolicity \cite{pe72}, equivalent to Leray's original
definition \cite{le52}. The strong causality of $(\mathcal{V},g)$
 follows if we prove that each slice
$\mathcal{M}_{t}$ intersects more than once  no inextendible,
future-directed causal curve $C:\mathcal{I}\rightarrow
\mathcal{V}:s \mapsto C(s)$. If $n=(-N,0)$ is the timelike normal
to $\mathcal{M}_{t}$, the
 tangent to this curve is such that,
\begin{equation}
g\left(\frac{dC}{ds},n\right)\equiv -N\frac{dt}{ds}<0,
\end{equation}
therefore on $C$ we have that,
\begin{equation}
\frac{dt}{ds }>0,
\end{equation}
and hence $C$ can be reparametrized using $t$ and cuts each
$\mathcal{M}_{t}$ at most once. To prove that the sets of the form
$J^+(u)\bigcap J^-(v)$ with $u,v\in\mathcal{V}$ are compact we
proceed as follows. Suppose there is a pair of points
$(x_{1},t_{1}), (x_{2},T)$ of $\mathcal{V}$ such that the set
$J^+((x_{1},t_{1}))\bigcap J^-((x_{2},T))$ is noncompact. This
means that there exists a future-directed, causal curve $C$ from
$(x_{1},t_{1})$ to $ (x_{2},T)$ which is inextendible. Consider a
Cauchy sequence of numbers $(t_{n})$ which converges to $T$ and
the corresponding points ($c_{n},t_{n})$ of the curve $C,$ where
$c_{n}$ (with components $ C^{i}(t_{n})$) are points of
$\mathcal{M}$. It follows that the sequence  $c_{n}$ cannot
converge to the point $c(T)$. But this is impossible, since the
estimates of \cite{ch-co02}, p. 347, show that  $c_{n}$ is a
Cauchy sequence in the complete Riemannian manifold
$(\mathcal{M},\gamma)$. Thus the sets $J^+(u)\bigcap J^-(v)$ are
compact and hence $(\mathcal{V},g)$ is globally hyperbolic.

$(2)\Rightarrow (1).$ Suppose that $(\mathcal{M}_{0},\gamma )$ is
not complete. Then from the Hopf-Rinow theorem  we can find a
geodesic $c:[0,\delta)\rightarrow\mathcal{M}_{0}$ of finite length
which cannot be extended to the arclength value $s=\delta
<\infty$. We take two times $t_{1}< t_{2}$ greater than zero, such that
$\delta<(t_{2}-t_{1})/2\equiv\delta^{*}/2 $. Since $\delta$ is
given by the geometry of the slice, this is a hypothesis on
$t_{2}-t_{1}$, i.e., on the minimum length of the spacetime slab.

Define on $\mathcal{V}$ the future-directed causal curve
$\bar{c}:[0,\delta )\rightarrow\mathcal{V}$ with
$$\bar{c}=(t+t_{1}=s,c(s)),$$  and the past-directed causal curve
$\tilde{c}:[0,\delta )\rightarrow\mathcal{V}$ with
$$\tilde{c}=(\delta^{*}-t+t_{1}=s,c(s)).$$ The curve $\bar{c}$ is
causal if
\begin{equation}\label{causal cond}
-N^{2}(\bar{c})+g_{ij}(\bar{c})\left(\frac{dc^{i}}{ds}+\beta^{i}\right)
\left(\frac{dc^{j}}{ds}+\beta^{j}\right)\leq 0.
\end{equation}
Since $c$ is a geodesic on $(\mathcal{M}_{0},\gamma )$ we have
\begin{equation}
\gamma _{ij}(c)\frac{dc^{i}}{ds}\frac{dc^{j}}{ds}=1,
\end{equation}
and therefore Condition (\ref{causal cond})  will hold
whenever\footnote{inequality (\ref{causal cond2}) could be lifted
by replacing $\bar{c}$ by a curve $(k(t+t_{1})=s,c(s)),$ $0\leq
s<\delta $ with $k$ an appropriate positive number. This curve
$\bar{c}$ is in the past of $c\equiv (t=0,s=c(s)),$ $0\leq
s<\delta $ if on it $t<0,$ i.e. $t_{1}>k^{-1}\delta .$ Analogous
reasoning for $\tilde{c}.$ }
\begin{equation}\label{causal cond2}
D+B\leq N_{m}^2.
\end{equation}
Similar reasoning for the curve $\tilde{c}$.

The curve $\bar{c}$ starts from the point $(-t_{1},c(0))$ and
proceeds to the future in the past of $c$, while $\tilde{c}$
starts from the point $(t_{2},c(0))$ and develops to the past in
the future of $c\,(\equiv (t=0,c(s)) )$. Therefore for each $t\in
[0,\delta )$, since $t<\delta^{*}-t$,
 we conclude that
\begin{equation}
(-t_{1},c(0))\prec\bar{c}(t)\ll\tilde{c}(t)\prec(t_{2},c(0)),
\end{equation}
where $\prec, \ll$ are the causality ($J$) and  chronology ($I$)
relations respectively. It follows that the diamond-shaped set
$J^{+}(-t_{1},c(0))\bigcap J^{-}(t_{2},c(0))$ contains the curve
$\bar{c}([0,\delta ))$. But since the set $c([0,\delta ))$ does
not have compact closure in $(\mathcal{M}_{0},\gamma)$, it follows
that $\bar{c}([0,\delta ))$ cannot have compact closure in
$J^{+}(-t_{1},c(0))\bigcap J^{-}(t_{2},c(0))$. This is however
impossible, for the set $J^{+}(-t_{1},c(0))\bigcap
J^{-}(t_{2},c(0))$ is compact because $\mathcal{V}$ is globally
hyperbolic. Hence, the curve $c$ is extendible.
\end{proof}

\section{Geodesic completeness of trivially sliced spaces}
\noindent In this Section we are interested in the question under
what conditions is global hyperbolicity equivalent to geodesic
completeness. What is the class of sliced spaces in which such an
equivalence holds? In a sliced space belonging to this class, in
view of the results of the previous Section, geodesic completeness
of the spacetime would be guessed very simply: It would suffice to
look at the completeness of a slice.

The tangent vector $u$ to a geodesic parametrized by arc length,
or by the canonical parameter in the case of a null geodesic, with
components $dx^{\alpha}/ds$ in the natural frame, satisfies in an
arbitrary frame the geodesic equations,
\begin{equation}
u^{\alpha }\nabla _{\alpha }u^{\beta }\equiv u^{\alpha }\partial
_{\alpha }u^{\beta }+\omega _{\alpha \gamma }^{\beta }u^{\alpha
}u^{\gamma }=0. \label{geo-eqn}
\end{equation}
In the adapted frame the components of $u$ become,
\begin{equation}
u^{0}=\frac{dt}{ds},\quad u^{i}=\frac{dx^{i}}{ds}+\beta
^{i}\frac{dt}{ds},
\end{equation}
while the Pfaff derivatives are given by,
\begin{equation}
\partial_{0}\equiv\partial_{t}-\beta^{i}\partial_{i}, \quad\partial_{i}\equiv%
\frac{\partial}{\partial x^{i}}.
\end{equation}
It holds therefore that,
\begin{equation}
u^{\alpha}\partial_{\alpha }u^{\beta }\equiv \frac{dt}{ds}
\left(\partial_{t}-\beta^{i}\partial_{i}\right)u^{\beta}
+\left(\frac{dx^{i}
}{ds}+\beta^{i}\frac{dt}{ds}\right)\partial_{i}
u^{\beta}\equiv\frac{ du^{\beta }}{ds}.
\end{equation}
Since $u^{0}\equiv dt/ds$, setting
\begin{equation}
v^{i}=\frac{dx^{i}}{dt}+\beta ^{i},
\end{equation}
so that
\begin{equation}
u^{i}=v^{i}u^{0},\label{change}
\end{equation}
 Eq. (\ref{geo-eqn}) with $\beta =0$ gives the 0-component of the
 geodesic equations which can be written in the
form,
\begin{equation}
\partial_{t}u^0+u^0\left(\omega _{00}^{0}+2\omega
_{0i}^{0}v^{i}+\omega _{ij}^{0}v^{i}v^{j}\right)=0. \label{3.7}
\end{equation}
On the other hand, the $k$-component of the geodesic equations is
\begin{equation}
\partial_{t}v^{k}+v^{i}\partial_{i}v^{k}+\omega
_{00}^{k}+2\omega _{0i}^{k}v^{i}+\omega _{ij}^{k}v^{i}v^{j} =0.
\label{k-comp}
\end{equation}
Using the expressions for the connection coefficients, we conclude
that when the lapse and shift are constant functions and
$\partial_{0}g_{ij}=0$, Eq. (\ref{3.7}) gives
\begin{equation}
t=\mathrm{const.}\times s,
\end{equation}
while Eq. (\ref{k-comp}), using (\ref{change}), becomes,
\begin{equation}
\frac{d^{2}x^{i}}{ds^2}+\tilde{\Gamma}^{k}_{ij}\frac{dx^{i}}{ds}\frac{dx^{j}}{ds}=0,
\end{equation}
that is the geodesic equation in the Riemannian manifold
$\mathcal{M}$. This result means that  in a sliced space
$(\mathcal{V},g)$ with constant lapse and shift and
time-independent spatial metric $g_{ij}$, called here a
\emph{trivially sliced space}, a curve
$x^{\alpha}(s)=(t(s),x^{i}(s))$ is a geodesic if and only if
$x^{i}(s)$ is a geodesic in the Riemannian manifold $\mathcal{M}$.

From the result proved above, it follows that  all geodesics of
trivially sliced
$(\mathcal{V},g)\equiv\mathcal{M}\times\mathbb{R}$ (with $g\equiv
-dt^2 +g_{ij}(x)dx^{i}dx^{j}$) have one of the following forms:
either $(\mu , x_{0})$ for some $x_{0}\in\mathcal{M}_{0}$ and
$\mu$ a constant, or $(\mu ,c(s))$ with $c(s)$ a geodesic of
$\mathcal{M}_{0}$, or $(\mu s,x_{0})$, or in the general case
$(\mu s,c(s))$ (we have taken $t=\mu s$).

Evidently, such geodesics will be complete if and only if $c(s)$
is complete,  hence  $(\mathcal{V},g)$ is timelike and null
geodesically complete, if and only if $(\mathcal{M}_{0},\gamma)$
is complete. We have therefore the following result.
\begin{theorem}
Let  $(\mathcal{V},g)$ be a trivially sliced space. Then the
following are equivalent:
\begin{enumerate}
\item The spacetime $(\mathcal{V},g)$ is timelike and null geodesically complete
\item $(\mathcal{M}_{0},\gamma )$ is a complete Riemannian manifold
\item The spacetime $(\mathcal{V},g)$ is globally hyperbolic.
\end{enumerate}
\end{theorem}
This result provides a partial converse to the completeness
theorem given in \cite{ch-co02} (Thm. 3.2) under the restricted
assumptions given above, and gives necessary and sufficient
conditions for the nonexistence of singularities in this case.

It appears that to prove a generic singularity theorem for  more
general sliced spaces having non-constant lapse and shift
functions and time-dependent spatial metric, one needs information
about the extrinsic curvature of the slices.

\section*{Acknowledgement}
I am  indebted to Y. Choquet-Bruhat for her precious comments and
critisism.

\end{document}